 \documentclass[pmlr,twocolumn]{jmlr} 

\usepackage{booktabs}
\usepackage[load-configurations=version-1]{siunitx} 

\usepackage{makecell}

\theorembodyfont{\upshape}
\theoremheaderfont{\scshape}
\theorempostheader{:}
\theoremsep{\newline}

\jmlrvolume{ML4H Extended Abstract Arxiv Index}
\jmlryear{2020}
\jmlrsubmitted{2020}
\jmlrpublished{}
\jmlrworkshop{Machine Learning for Health (ML4H) 2020} 

\title[Detection of missed pneumothorax on chest x-rays]{Pneumothorax and chest tube classification on chest x-rays for detection of missed pneumothorax}

  \author{%
  \Name{Benedikt Graf} \Email{ben.graf@ibm.com}\\
  \Name{Arkadiusz Sitek} \Email{arek@ibm.com}\\
  \Name{Amin Katouzian} \Email{akatouz@us.ibm.com}\\
  \Name{Yen-Fu Lu} \Email{yen-fu.luo@ibm.com}\\
  \Name{Arun Krishnan} \Email{arun.krishnan100@ibm.com}\\
  \addr IBM Watson Health Imaging
    \AND
    \Name{Justin Rafael} \Email{jrafael@rasf.net}\\
    \addr Baptist Health South Florida, Department of Radiology
    \AND
    \Name{Kirstin Small} \Email{ksmall1@bwh.harvard.edu}\\
    \addr Brigham and Women's Hospital, Department of Radiology
    \AND
    \Name{Yiting Xie} \Email{yiting.xie@ibm.com}\\
    \addr IBM Watson Health Imaging
  }

\begin{document}

\maketitle

\begin{abstract}
Chest x-ray imaging is widely used for the diagnosis of pneumothorax and there has been significant interest in developing automated methods to assist in image interpretation. We present an image classification pipeline which detects pneumothorax as well as the various types of chest tubes that are commonly used to treat pneumothorax. Our multi-stage algorithm is based on lung segmentation followed by pneumothorax classification, including classification of patches that are most likely to contain pneumothorax. This algorithm achieves state of the art performance for pneumothorax classification on an open-source benchmark dataset. Unlike previous work, this algorithm shows comparable performance on data with and without chest tubes and thus has an improved clinical utility. To evaluate these algorithms in a realistic clinical scenario, we demonstrate the ability to identify real cases of missed pneumothorax in a large dataset of chest x-ray studies.
\end{abstract}

\begin{figure}[htbp]
\floatconts
  {fig:Chest-tubes}
  {\caption{Chest X-ray with a chest tube}}
  {\includegraphics[width=.85\linewidth]{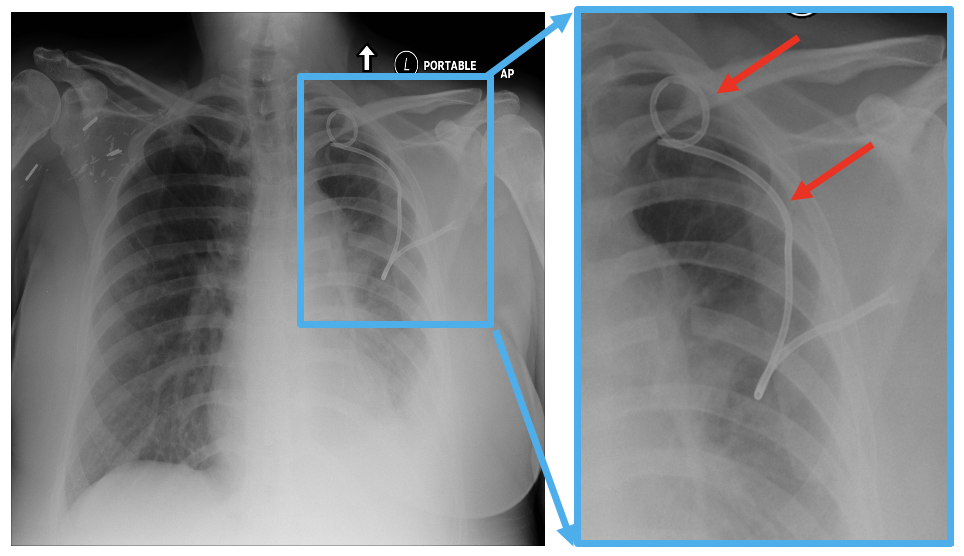}}
\end{figure}

\begin{figure*}[htbp]
\floatconts
  {fig:diagram}
  {\caption{System for detection of missed pneumothorax on chest x-rays. The image classification pipeline (red box) detects both pneumothorax and chest tubes while the NLP algorithm detects positive mentions of pneumothorax in the associated radiology report. The system identifies studies without chest tubes that potentially have a missed pneumothorax.}}
  {\includegraphics[width=.85\linewidth]{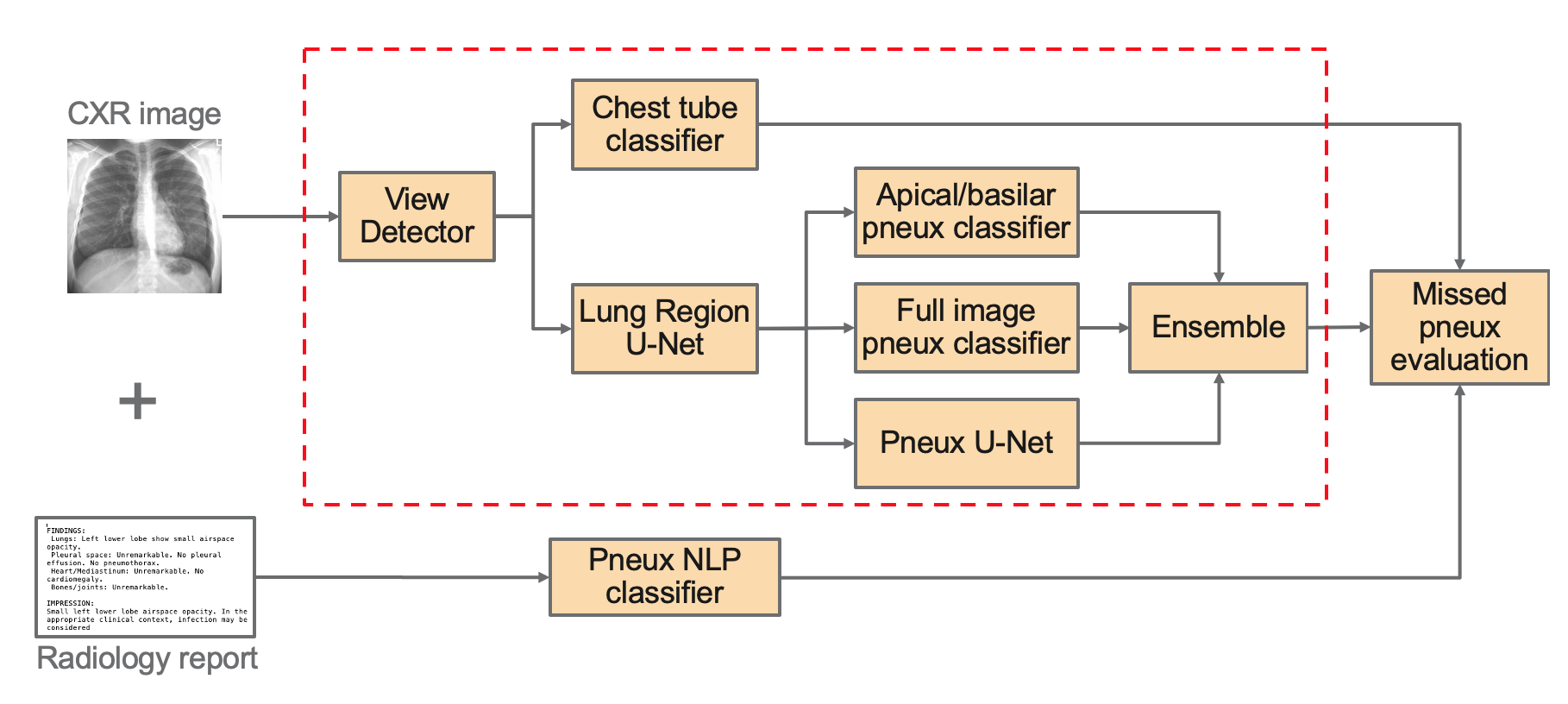}}
\end{figure*}

\section{Introduction}
\label{sec:intro}

Pneumothorax is a potentially life-threatening condition where air collects in the pleural space between the lung and chest wall. There has been significant interest in developing automated methods for detecting pneumothorax on chest x-rays. One notable limitation of previous work is the lack of associated methods for chest tube recognition. Chest tubes are flexible plastic tubes that are inserted through the chest wall to drain air from the pleural space in order to treat pneumothorax. Significant proportions of pneumothorax cases in chest x-ray datasets contain chest tubes ($>$70\% in \citep{Majkowska2020}). Since chest tubes may appear as prominent features on chest x ray images (see \figureref{fig:Chest-tubes}), there is a risk that deep learning models will detect the presence of chest tubes rather than the pneumothorax itself. Most prior work does not consider the effect of chest tubes \citep{Irvin2019,tCheXNet,Rajpurkar2017,Wang2017,Taylor2018} and the recent work from \citep{Majkowska2020} demonstrates a 13.2\% drop in sensitivity when evaluating data without chest tubes.

\begin{table*}[hbtp]
\floatconts
  {tab:results}
  {\caption{Results for pneumothorax classification. Best values for each column are in bold.}}
  {\begin{tabular}{lcccc}
  \toprule
  \bfseries \makecell{Method} & \bfseries \makecell{AUC \\ (all data)} & \bfseries \makecell{AUC \\ (no tubes)} & \makecell{AUC \\ (only tubes)} & \makecell{AUC \% change \\ with no tubes}\\
  \midrule
  \cite{Rajpurkar2017} & 0.895 & 0.816 & 0.914 & -8.8\% \\
  \cite{tCheXNet}  & 0.844 & 0.787 & 0.858 & -6.8\% \\
  \cite{Majkowska2020}  & 0.940 & 0.890 & - & -5.3\% \\
  A - DN full image         & 0.941 & 0.941 & 0.941 & 0.0\% \\
  B - DN apical/basilar    & 0.932 & 0.878 & 0.946 & -5.8\% \\
  C - U-Net                 & 0.921 & 0.927 & 0.919 & \textbf{0.7}\% \\
  Ensemble A + C            & 0.952 & \textbf{0.953} & 0.952 & 0.1\% \\
  Ensemble A + B + C        & \textbf{0.958} & 0.948 & \textbf{0.960} & -1.0\% \\

  \bottomrule
  \end{tabular}}
\end{table*}

\section{Pneumothorax and chest tube classification pipeline}
\label{sec:classification}

In this work we present an image classification pipeline that detects both pneumothorax and chest tubes. The pipeline is shown in the red box in \figureref{fig:diagram}. All images are first evaluated by an X-ray view classification deep learning model so that only frontal (AP or PA) images are processed by the rest of the pipeline. Pneumothorax classification consists of first performing lung segmentation with a U-Net \citep{Ronneberger2015}. This is followed by three separate approaches for pneumothorax classification which are ensembled. The first method crops the image to the lung region and then classifies the image with a DenseNet-121 model \citep{Huang2016}. The second method utilizes a U-Net to segment the pneumothorax region on the same cropped and resized image. The third method extracts four patches from the left/right, basilar/apical lung regions. These specific lung regions are chosen as they are the most likely to contain a small pneumothorax. The three different models are ensenbled to create the final output for pneumothorax classification. The chest tube classification algorithm is a DenseNet-121 model. The input to this model is the full chest X-ray image without any cropping. 

All deep learning models were trained separately on both open-source data and a private multi-site data set with annotations done by radiologists. For pneumothorax segmentation, we used 8,222 images from the NIH ChestXray14 dataset \citep{Wang2017} with pixel level annotations that were provided as part of the Kaggle Pneumothorax Challenge \citep{kaggle}. We also used an additional 6,011 images from the private dataset.  The pneumothorax and chest tube classification algorithms were developed using 25,173 images from the private dataset. Training labels for pneumothorax and chest tubes were generated in a semi-automated way from the radiology reports. These sentences were manually categorized as positive or negative to generate verified image level labels. When available, the location of the pneumothorax (left/right and base/apex) was also captured from the report for training the apical/basilar classifier. Additional details about the training data can  be found in \appendixref{apd:first}.

\begin{figure*}[htbp]
\floatconts
  {fig:missed-finding}
  {\caption{(a) Example of a missed pneumothorax detected by the image and NLP algorithms, with red arrows highlighting the edge of the collapsed lung. (b) Findings and impression section of the corresponding radiology report demonstrating that the pneumothorax was not reported.}}
  {\includegraphics[width=.9\linewidth]{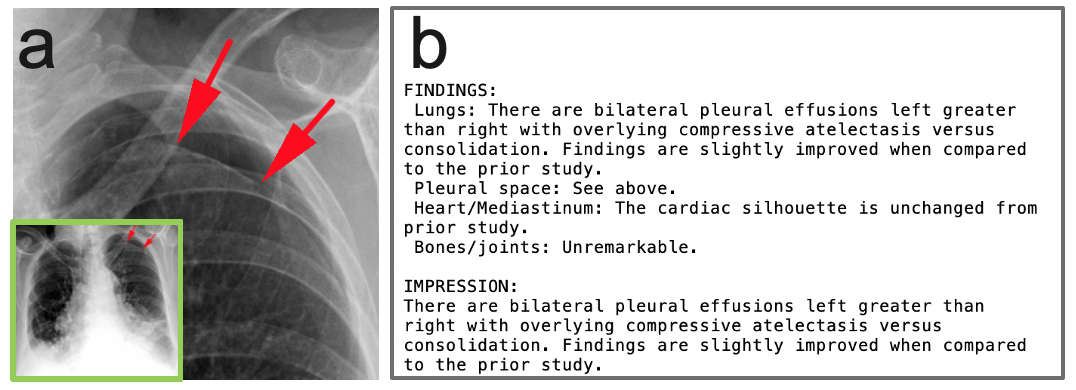}}
\end{figure*}

\section{Missed pneumothorax detection}
\label{sec:missed-pneux}

To evaluate these algorithms in a realistic clinical scenario, we combine the image algorithms with natural language processing (NLP) to identify studies with potential missed pneumothorax (pneumothorax detected on images but not mentioned in radiology report).  Missed findings that are clearly noticeable on retrospect are a known problem in radiology that can result from a number of factors such as reader fatigue and distractions such as phone calls \citep{vanGinneken2011,Brady2017}. The idea of combining image classification algorithms with NLP to detect missed findings was recently demonstrated for intracranial hemorrhage on non-contrast CT scan \citep{Rao2020}.  

The NLP (previously described in \cite{Secret2017}) analyzes radiology reports and classifies the report as positive if there is any positive mention of pneumothorax. Any study that has a negative output from the NLP, a negative output for chest tube image classification and a positive output for pneumothorax image classification are identified as potentially having a missed pneumothorax. Studies with a chest tube are not considered for potential missed findings since the presence of the chest tube indicates that there is already a clinical awareness of a pneumothorax or other pleural space abnormality.

\section{Results}

We evaluated pneumothorax classification performance using the radiologist adjudicated test set made available by \cite{Majkowska2020} (see \appendixref{apd:first}).  We also compare our results with two other models that have been open sourced \citep{tCheXNet,Rajpurkar2017}. These results are shown in \tableref{tab:results}. In addition to having a higher overall AUC, our model maintains similar performance for samples that do not contain chest tubes. In comparison, the other models we evaluated had a significant drop in AUC (5.3\%-8.8\%). 

We also evaluated the chest tube classification algorithm on a test set of 2,977 images with radiologist annotated ground truth. AUC was .977 for the classification of standard chest tubes and .987 for pigtail catheters. 

To demonstrate the ability to identify true missed findings in a realistic clinical setting, we analyzed a set of 20,000 chest x-ray studies and their associated radiology reports. Using the combined image and NLP algorithm pipeline, we identified 62 studies as having a potential missed pneumothorax. A board-certified radiologist reviewed both the radiology report text and images and identified nine cases of missed pneumothorax (see  example in \figureref{fig:missed-finding}).  

\section{Summary of contributions}

\begin{itemize}
  \item We developed an image classification pipeline that achieves state of the art performance for pneumothorax classification and also includes the ability to identify chest tubes.
  \item We demonstrate that, unlike previous works, this pipeline achieves comparable performance on pneumothorax classification in data with and without the presence of chest tubes which significantly improves the clinical utility.
  \item We utilize these image classification algorithms in combination with an NLP algorithm to demonstrate the ability to identify real cases of missed pneumothorax in a large and diverse dataset of chest X-ray studies
\end{itemize}

\bibliography{jmlr-sample}

\appendix

\section{Data Details}\label{apd:first}

\tableref{tab:training-datasets} summarizes the data used for development of the algorithms and \tableref{tab:test-datasets} summarizes the data used for testing and evaluation.

\begin{table*}[hbtp]
\floatconts
  {tab:training-datasets}
  {\caption{Datasets used for development}}
  {\begin{tabular}{lccl}
  \toprule
  \bfseries Dataset description & \bfseries \makecell{Total \# \\ samples} & \bfseries \makecell{\# positive\\ samples} & \bfseries Ground truth \\
  \midrule
  \makecell{NIH ChestX-ray14 \\and SIIM-Kaggle} & 8,222 & 1,059 & \makecell{Pixel level \\ annotation}\\
  \makecell{Pneux segmentation} & 6,011 & 1,958 & \makecell{Pixel level \\ annotation}\\ 
  \makecell{Pneux and chest \\ tube classification} & 25,173 & \makecell{6,597 pneux full image \\ 3,217 pneux apex/base \\ 5,742 chest tube} & \makecell{Image level labels\\ from report}\\
  \bottomrule
  \end{tabular}}
\end{table*}

\begin{table*}[hbtp]
\floatconts
  {tab:test-datasets}
  {\caption{Datasets used for testing and evaluation}}
  {\begin{tabular}{lccl}
  \toprule
  \bfseries Dataset description & \bfseries \makecell{Total \# \\ samples} & \bfseries \makecell{\# positive\\ samples} \\
  \midrule
  Chest tube classification & 2,977 & 823 (506 standard, 329 pigtail) \\
  Pneumothorax classification & 1,962 & \makecell{195 (156 with chest tubes, 39 without)}\\ 
  \makecell{Missed pneumothorax evaluation} & 20,000 & \makecell{NA}\\
  \bottomrule
  \end{tabular}}
\end{table*}

\end{document}